\documentstyle[psbox]{elsart}

\begin{document}
\begin{frontmatter}
\title{Subpixel Spatial Resolution of the X-Ray Charge-Coupled Device
Based on the Charge Cloud Shape}
 \date{\today}
 %received: September 1, 2000
 accepted December, 14, 2000
 to {\it Jpn. J. Appl. Phys.}
\date{}

\author[osaka]{J. HIRAGA}
\author[osaka]{H.TSUNEMI}
\author[osaka]{E.MIYATA}

\address[osaka]{Department of Earth and Space Science, Graduate School
of Science, Osaka University, 1-1 Machikaneyama-cho, Toyonaka, Osaka
5600043, Japan \\
CREST, Japan Science and Technology Corporation (JST)\\
}

\begin{abstract}

When an X-ray photon lands into a pixel (event pixel), the primary
charge is mainly collected into the event pixel.  If the X-ray landing
position is sufficiently close to the pixel boundary, the primary charge
spills over to the adjacent pixel forming split events.  We can easily
understand that there are three parameters coupled together; the X-ray
landing position inside the pixel, the X-ray event pattern and the
primary charge cloud shape.  We can determine any one of them from the
other two parameters.  Since we know the charge cloud shape using the
multi-pitch mesh experiment, we can calculate the X-ray landing position
with subpixel resolution using the event pattern. We applied our method
to Ti-K X-rays for the charge-coupled device with $12\,\mu$m square
pixel.  Once the primary charge splits into the adjacent pixel, we can
determine the X-ray landing position with subpixel resolution.  Using
three- or four-pixel split events, we can determine the X-ray landing
position with an accuracy of less than $1\,\mu$m.  For a two-pixel split
event, we obtained a similar position accuracy in the split direction
with no improvement in the direction perpendicular to it.  We will
discuss the type of CCD which can achieve the subpixel resolution for
the entire area of the CCD.

\end{abstract}

\begin{keyword}
 charge-coupled device, X-ray event, split event, subpixel spatial
 resolution
\end{keyword}

\end{frontmatter}

\section{Introduction}

A charge-coupled device (CCD) is widely used in optical imaging,
particularly in the digital camera for commercial use.  It mainly
employs an interline type CCD. The pixel size is approximately a few $\mu$m square which produces a few-mega-pixel image~\cite{watanabe00}~.
Although the interline type CCD has a relatively poor opening area, the
micro-lens array system can increase the effective opening area up to
70\% or higher. Furthermore, the CCD for optical region requires the
depletion region to be at the most several $\mu$m.  However, the CCD
for X-ray use has different characteristics from that for optical use.
It has a relatively good spatial resolution and a moderate energy
resolution.  Since the micro-lens array does not work for X-ray photons,
the frame transfer type CCD is required for X-ray use to have a high
opening area.  The low energy X-ray, especially below $1\,$keV, is
easily photoabsorbed, which requires the material above the depletion
region to be as thin as possible. The cross-over of the gate
structure above the depletion region makes the low energy X-ray
responsivity within the pixel complicated.  The photoabsorption depth in
silicon for $10\,$keV X-ray is about $100\,\mu$m, which requires a much
thicker depletion region for X-ray use than that for optical use.  

Several CCDs for X-ray use, particularly for X-ray astronomy, have been
developed to date.  The ASCA satellite launched in February, 1993,
employs CCDs with $27\,\mu$m square pixels with a depletion depth of
$35\,\mu$m~\cite{tanaka94,burke94}.  The Chandra Observatory launched in
July, 1999, employs CCDs with $24\,\mu$m square pixels with a depletion
depth of about $70\,\mu$m~\cite{chandra,burke97}.  The XMM-Newton
Observatory launched in December, 1999, employs two types of
CCDs~\cite{XMM1} : one has a $150\,\mu$m square pixel with a depletion
depth of $280\,\mu$m~\cite{PNCCD} while the other has a $40\,\mu$m
square pixel with a depletion depth of $40\,\mu$m~\cite{MOSCCD}.  These
CCDs for X-ray use are developed to have thick depletion regions rather
than to have high spatial resolution.

When an X-ray photon is photoabsorbed in a CCD, it generates a number of
electrons proportional to the incident X-ray energy.  The electrons
generated by X-ray photons usually expand by the diffusion process while
they are pulled to the potential well of the CCD pixel. They usually
form a charge cloud of a finite size.  They are collected into several
pixels forming various types of event pattern (\lq grade\rq) depending
on how they split.  When the entire charge is collected into one pixel
(no surrounding pixels have a signal), it is called a \lq single pixel
event\rq: when it splits into more than one pixel, it is called a \lq
split pixel event\rq.  The event grade is determined both by the landing
position of X-rays within the pixel and by the charge cloud shape.  The
X-ray responsivity depends on the event grade as well as its landing
position.  This is quite in contrast to the optical photon that
generates only one or two electrons.  Therefore, there is no concept of
the event grade in an optical region.

 We introduced a new technique, \lq mesh experiment\rq, which enables us
 to measure the X-ray responsivity of the CCD with subpixel
 resolution~\cite{tsunemi97}.  In this experiment, a parallel X-ray beam
 is irradiated onto the CCD chip above which a metal mesh is placed.
 The mesh has many small holes which are periodically spaced.  There are
 two types of experiments; a single pitch mesh experiment and a
 multi-pitch mesh experiment.  If the small hole spacing on the mesh is
 equal to the CCD pixel size, it is called a single pitch mesh
 experiment and if it is a multiple of the CCD pixel size, it is called
 a multi-pitch mesh experiment.  Using this technique, the detailed gate
 structures were directly measured from the X-ray absorption
 feature~\cite{pivovaroff98,yoshita98}.  The details of the mesh
 experiment are described in literature by Tsunemi et
 al.~\cite{tsunemi98}

 Using the multi-pitch mesh experiment, we can unequivocally identify
 the X-ray landing position inside the pixel.  In this way, we can
 directly measure the event grade, and how the charge cloud splits into
 surrounding pixels, as a function of the landing position inside the
 pixel.  We can find out that the fraction of the charge splitting into
 the adjacent pixel increases as the X-ray landing position approaches
 the pixel boundary.  We can easily understand that there are three
 parameters coupled together; the X-ray landing position inside the
 pixel, the X-ray event grade and the charge cloud shape.  If we know
 two out of three parameters, we can calculate the third one.  The X-ray
 event grade is easily measured by the photon count CCD system.  The
 X-ray landing position can be directly measured by the mesh experiment,
 resulting us to measure the charge cloud shape~\cite{hiraga98}.  Once
 we know the charge cloud shape, we can measure the X-ray landing
 position without applying the mesh experiment.

In this paper, we briefly describe the mesh experiment and present the
method of determining the X-ray landing position for various types of
event grades in order to achieve the subpixel resolution.

\section{Experiment}

An X-ray photon photoabsorbed inside the CCD usually generates a number
of electrons, the primary charge cloud.  One X-ray photon photoabsorbed
generates one X-ray event, a series of pixels collecting a primary
charge cloud.  The primary charge cloud expands to some extent while it
is pulled to the potential well.  If the X-ray photon is photoabsorbed
below the depletion region, some fraction of the charge is collected as
a multi-pixel event.  If the X-ray photon is photoabsorbed in the
depletion region, the size of the primary charge cloud is at the 
most a few $\mu$m which is smaller than that of the CCD pixel size.
The actual charge cloud size depends on the photoabsorbed depth.
Therefore, we can distinguish the events photoabsorbed in the depletion
region from those photoabsorbed below the depletion region by the event
pattern.  The former shows the incident X-ray energy while the latter
does not due to the charge loss.  We will focus on the events
photoabsorbed in the depletion region.

We will obtain various grades of events depending on the X-ray landing
position.  A major fraction of the primary charge will be collected in
the pixel in which the X-ray lands.  We term this pixel as the \lq
event pixel\rq.  When the entire primary charge cloud is collected into
the event pixel, the surrounding pixels have no signal, resulting in a
single pixel event.  If the X-ray landing position in the event pixel is
sufficiently close to the pixel boundary, the primary charge splits into two
pixels forming a two-pixel split event.  If the X-ray landing position is
close to the event pixel corner, the primary charge splits into three or
four pixels forming three- or four-pixel split event or \lq corner
event\rq.

When we detect an X-ray event in the CCD, we usually see $3\times 3$
pixels as the event pattern.  The central pixel of the event pattern is
the event pixel that has the maximum charge in the 3 $\times$ 3 pixels.
The pixels around the event pixel are \lq surrounding pixels\rq.  In
practice, we introduce the split threshold level, $T_h$, in order to
test whether or not the pixel has charge.  If we find that the output
from the surrounding pixels exceeds $T_h$, we term the pixel as a \lq
split pixel\rq. In X-ray spectroscopy, we usually sum up the output from
the event pixel and those from the split pixels in order to evaluate the
incident X-ray energy.  Since each pixel usually contains noise level,
the best energy resolution is obtained by eliminating the pixels whose
output is less than $T_h$.  The overall noise level in our system is 10
electrons or less, therefore, we set $T_h$ to be 50 electrons so that we
do not treat the pixel with no signal as a split pixel by noise.

%%%%%%%%%%%% edited by jhiraga at 2000/10/23 %%%%%%%%%%%%%
We performed the mesh experiment employing a CCD (Hamamatsu Photonics,
N38 11-5A0N-2) consisting of $12\,\mu$m square pixels for X-ray use.
The mesh fabricated of gold has small holes of $2.1\,\mu$m diameter
spaced $48\,\mu$m apart. The pitch of mesh holes is four times larger
than the CCD pixel size which forms a multi-pitch mesh experiment.  The
mesh is placed just $1\,$mm above the CCD.  In this way, we can measure
what type of event grades are generated as a function of the X-ray
landing position inside the CCD pixel.

\section{Event Pattern}

When an X-ray photon lands on the CCD at $(X_{in}, Y_{in})$, the output, $D(n, m; X_{in}, Y_{in})$, from the $(n,m)$ pixel in the X-ray event is expressed as an integration of the primary charge cloud, $C(X, Y)$, over the pixel area that is given in eq.~(\ref{pixel_output}).
\begin{equation}
\label{pixel_output}
D(n, m; X_{in}, Y_{in}) = \int_{X_{n}}^{X_{n+1}}dX \int_{Y_{m 
}}^{Y_{m+1}}dY~ C(X-X_{in}, Y-Y_{in})
\end{equation}
where $X_{n}, X_{n+1}$, and $Y_{m}, Y_{m+1}$ represent the pixel
boundary of the X and Y coordinates, respectively.  Since the CCD
employed has a square shaped pixel, we set $X_{n+1} - X_{n} = Y_{m+1} -
Y_{m} \equiv L$, where $L$ represents the pixel size.  We assume that
all the CCD pixels are identical, therefore, we obtain the
relationship in eq.~(\ref{P_relation}) as below,
\begin{equation}
\label{P_relation}
%%D(n+k, m+\ell; X_{in}, Y_{in}) = D(n, m; X_{in}-k\,L, Y_{in}-\ell\,L)~.
D(n+k, m+l; X_{in}, Y_{in}) = D(n, m; X_{in}-k\,L, Y_{in}-l\,L)~.
\end{equation}

When the $(n,m)$ pixel represents the event pixel, the event pattern of
the $3\times 3$ pixels are expressed in the matrix below,
\begin{eqnarray}
\label{3x3pixels}
\left( 
\begin{array}{ccc}
D(n-1, m+1; X_{in}, Y_{in} )	& D(n, m+1; X_{in}, Y_{in})	& D(n+1, m+1;
X_{in}, Y_{in})	\\
D(n-1, m; X_{in}, Y_{in})	& D(n, m; X_{in}, Y_{in})	& D(n+1, m;
X_{in}, Y_{in})	\\
D(n-1, m-1; X_{in}, Y_{in})	& D(n, m-1; X_{in}, Y_{in})	& D(n+1, m-1;
X_{in}, Y_{in})	\\
\end{array} 
\right)
\end{eqnarray}
It can be rewritten in the form below using eq.~(\ref{P_relation}) as,
\begin{eqnarray}
\label{3x3Dn}
\left( 
\begin{array}{ccc}
D(n,m;X_{in}+L,Y_{in}-L)& D(n,m;X_{in},Y_{in}-L)& D(n,m;X_{in}-L,Y_{in}-L)\\
D(n,m;X_{in}+L,Y_{in})& D(n,m;X_{in},Y_{in})& D(n,m; X_{in}-L,Y_{in})\\
D(n,m;X_{in}+L,Y_{in}+L)& D(n,m;X_{in},Y_{in}+L)& D(n,m;X_{in}-L,Y_{in}+L)\\
\end{array} 
\right)
\end{eqnarray}

\begin{figure}[hbtn]
  \begin{center}
%   \begin{minipage}{1.0\textwidth}
    \psbox[xsize=0.45#1]{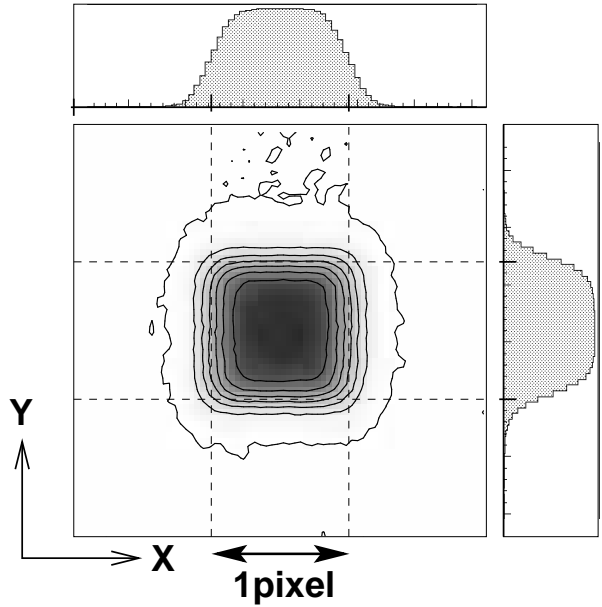} 
\vspace{-30mm}
 \caption{Signal output from the event pixel as a function of the X-ray landing position in the $3\times 3$ pixel region.  Projections along $X$ and $Y$ axes are also shown.}
   \label{Dn}
%   \end{minipage}
  \end{center}
 \end{figure}

In this way, the $3\times 3$ pixel data express the function of $D$.
Employing the multi-pitch mesh experiment, we can directly measure the
$3\times 3$ pixel data as a function of $(X_{in},Y_{in})$.  Then, we can
obtain the function, $D$, which is expressed as $D_n$ in the
literature~\cite{hiraga98}, from the raw data.  In practice, we relocate
the event pattern to match the function, $D$ in
eq.~(\ref{pixel_output}), which is shown in Fig.\,\ref{Dn} in $3\times
3$ pixel region.

\begin{figure}[hbtn]
  \begin{center}
    \psbox[r,xsize=0.4#1]{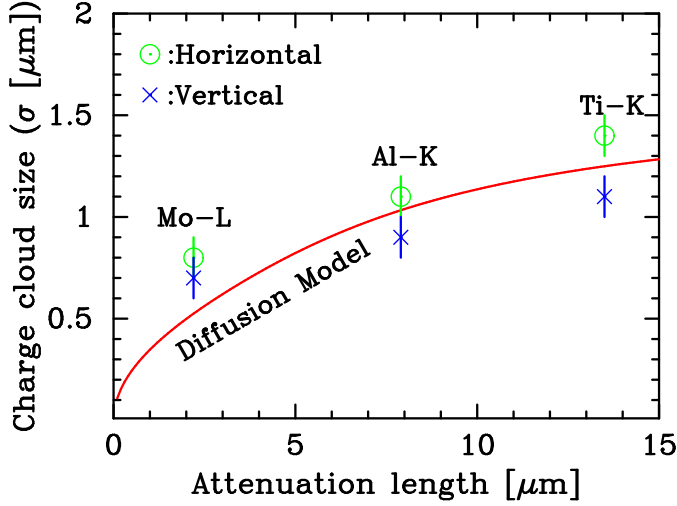} 
%\vspace{-30mm}
   \caption{Primary charge cloud size generated by various X-ray photons
 for Mo-L\,(2.3\,keV), Al-K\,(1.5\,keV) and Ti-K\,(4.5\,keV) as a function of
 the attenuation length in silicon.  The solid line
 represents the model calculation.}
   \label{charge_shape}
%   \end{minipage}
  \end{center}
 \end{figure}

Using the eq.~(\ref{pixel_output}), we find that the charge cloud shape
is obtained by differentiating $D$ with $(X_{in}, Y_{in})$.  The
detailed derivation of the charge cloud shape is given in the
literature~\cite{hiraga98}.  In the mesh experiment, we can confine the
X-ray landing position with the precision of the mesh hole. Therefore, the
differentiation of $D$ is a convolution between the charge cloud shape
and the mesh hole shape.  Since we know the mesh hole shape, we can
deduce the actual charge cloud shape which is well approximated by an
axial symmetric Gaussian function.  In this way, we can calculate the
primary charge cloud shape as a function of the incident X-ray energy as
shown in Fig.\,\ref{charge_shape}~\cite{tsunemi99}.

The longer the attenuation length of X-ray photons in silicon, the
bigger is the charge cloud size.  This is well approximated by a simple
diffusion model~\cite{hopkinson87}.  In the following analysis, we will
mainly focus on the Ti-K X-ray photons since it has the biggest charge
cloud size in our analysis.  Then, we will estimate the X-ray landing
position by comparing the event pattern with $D$.

\section{Data Reduction}

\subsection{Analysis method}

\begin{figure}[hbtn]
\begin{center}
 \psbox[xsize=0.45#1]{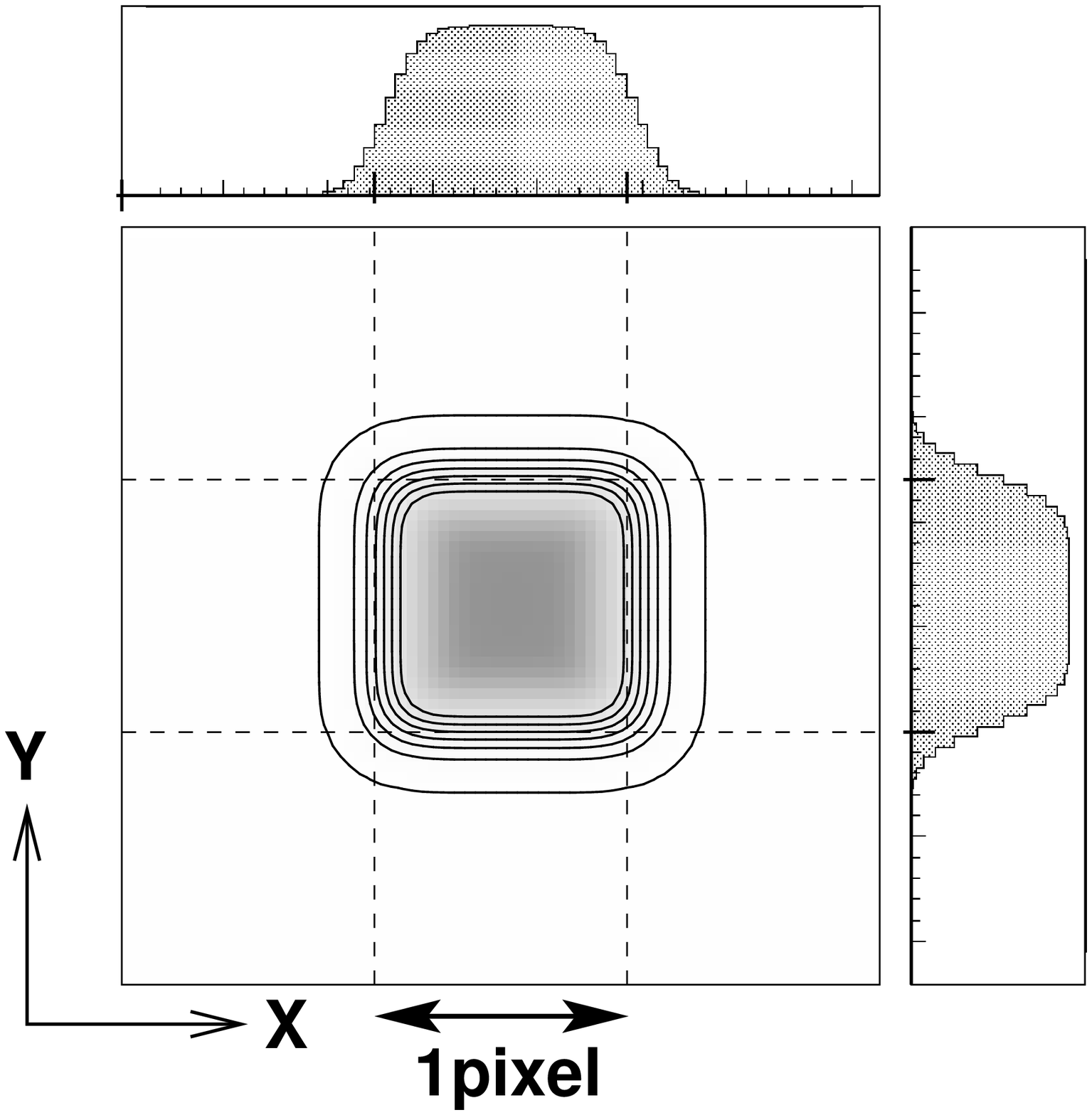}
\end{center}
\vspace{-30mm}
 \caption{Model calculation for $D$. The signal output from the event
 pixel is shown as a function of the X-ray landing position in the $3\times 3$
 pixel region.  This figure is free from the effect of the mesh hole
 shape.}
    \label{D_model}
 \end{figure}

Once we know the charge cloud shape, we need not employ the mesh any
more.  Since we can easily determine the  X-ray event pattern, we can calculate
the incident X-ray energy by adding up the signal contained in the
event.  Then we can estimate the charge cloud shape and calculate $D$
according to eq.~(\ref{pixel_output}).  Figure~\ref{D_model} shows
$D$ which is free from the effect of the mesh hole shape.

%\vspace*{-20mm}
%\marginpar{\fbox{Fig\ref{D_model}}}

When we obtain a $3\times 3$ event pattern, we can compare it with the
eq.~(\ref{3x3Dn}).  We calculate the sum of the square of differences
between the $3\times 3$ pixel data and $D$ as a function of
$(X_{in},Y_{in})$.  The estimated landing position is given as the
position when the sum is minimum.  Once we detect an X-ray event,
we always employ the $3\times 3$ pixel data regardless of its grade, whether
it is a single pixel event or not etc.  It should be noted that our method
does not take into account $T_h$.  However, we will sort the event by grades
to determine which grades can yield how much accuracy in estimating the
landing position.

\subsection{Estimated landing position}

Using this method, we can evaluate the X-ray landing position, $(X_{in},
Y_{in})$, while we know the location of the mesh hole shadow on the CCD,
$(X_{hole}, Y_{hole})$, in which the X-ray must have landed. Then the
difference, {\bf $\Delta$} $\equiv$ $(X_{in}, Y_{in}) - (X_{hole},
Y_{hole})$, shows the uncertainty of the position estimation.  We
calculate {\bf $\Delta$} for all X-ray events to estimate the position
accuracy.  Since the actual landing position of the X-ray photon is
somewhere inside the mesh hole, {\bf $\Delta$} is a convolution between
the mesh hole shape and the uncertainty of the estimation of the X-ray
landing position.

\begin{figure}[hbtn]
  \begin{center}
   \begin{minipage}{0.4\textwidth}
    \psbox[r,xsize=0.25#1]{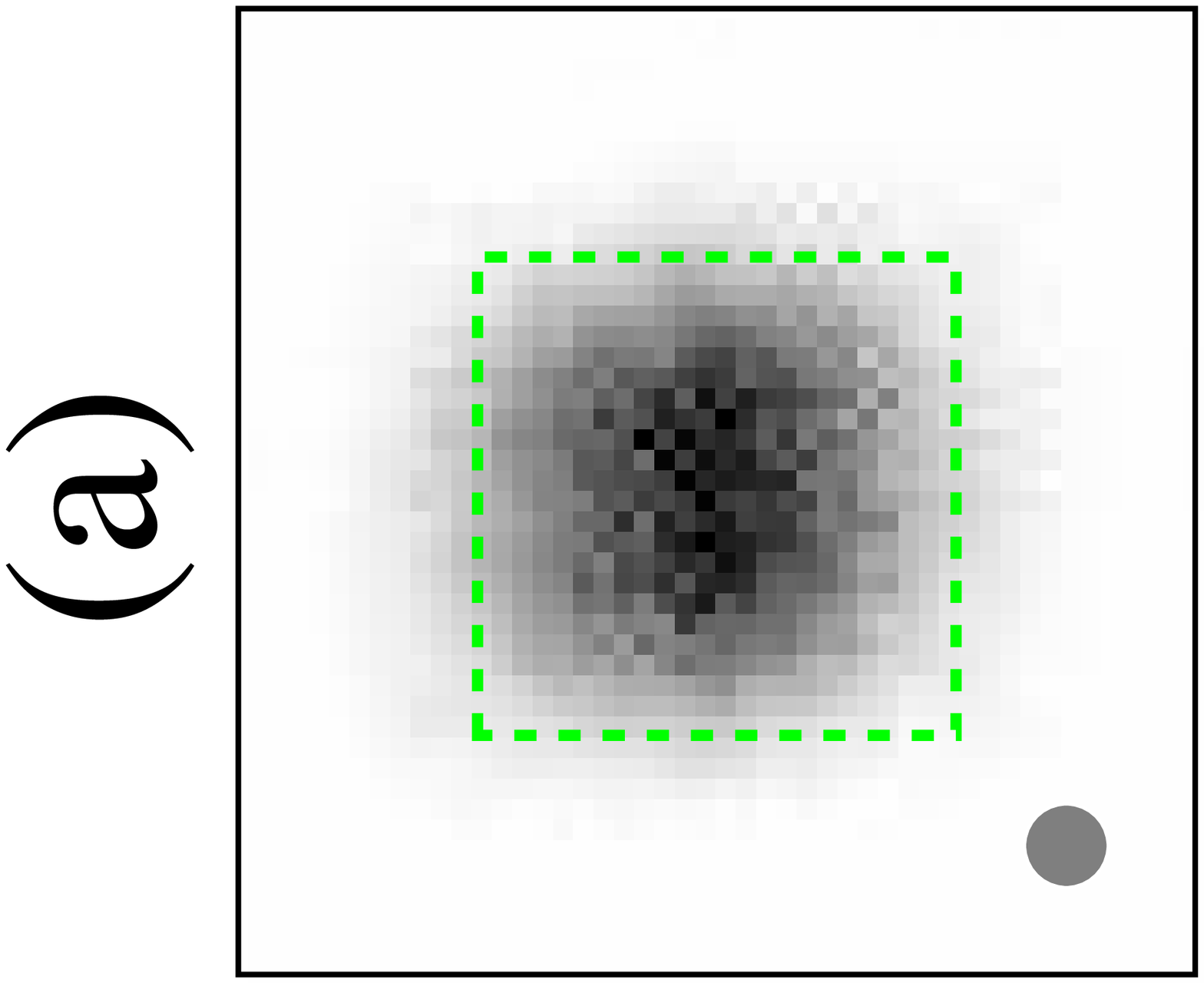} 
   \end{minipage}
   \begin{minipage}{0.4\textwidth}
    \psbox[r,xsize=0.25#1]{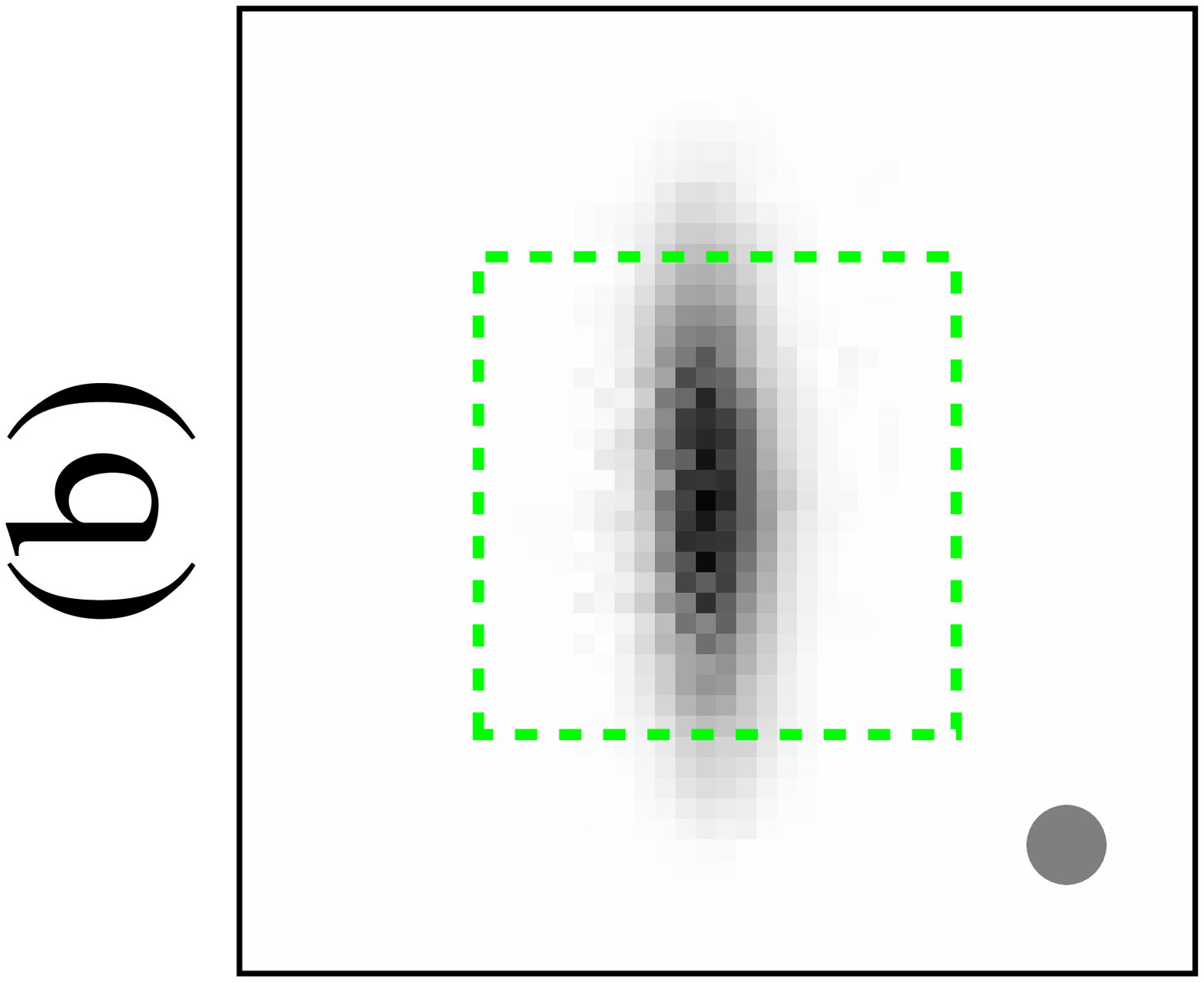} 
   \end{minipage}
   \begin{minipage}{0.4\textwidth}
    \psbox[r,xsize=0.25#1]{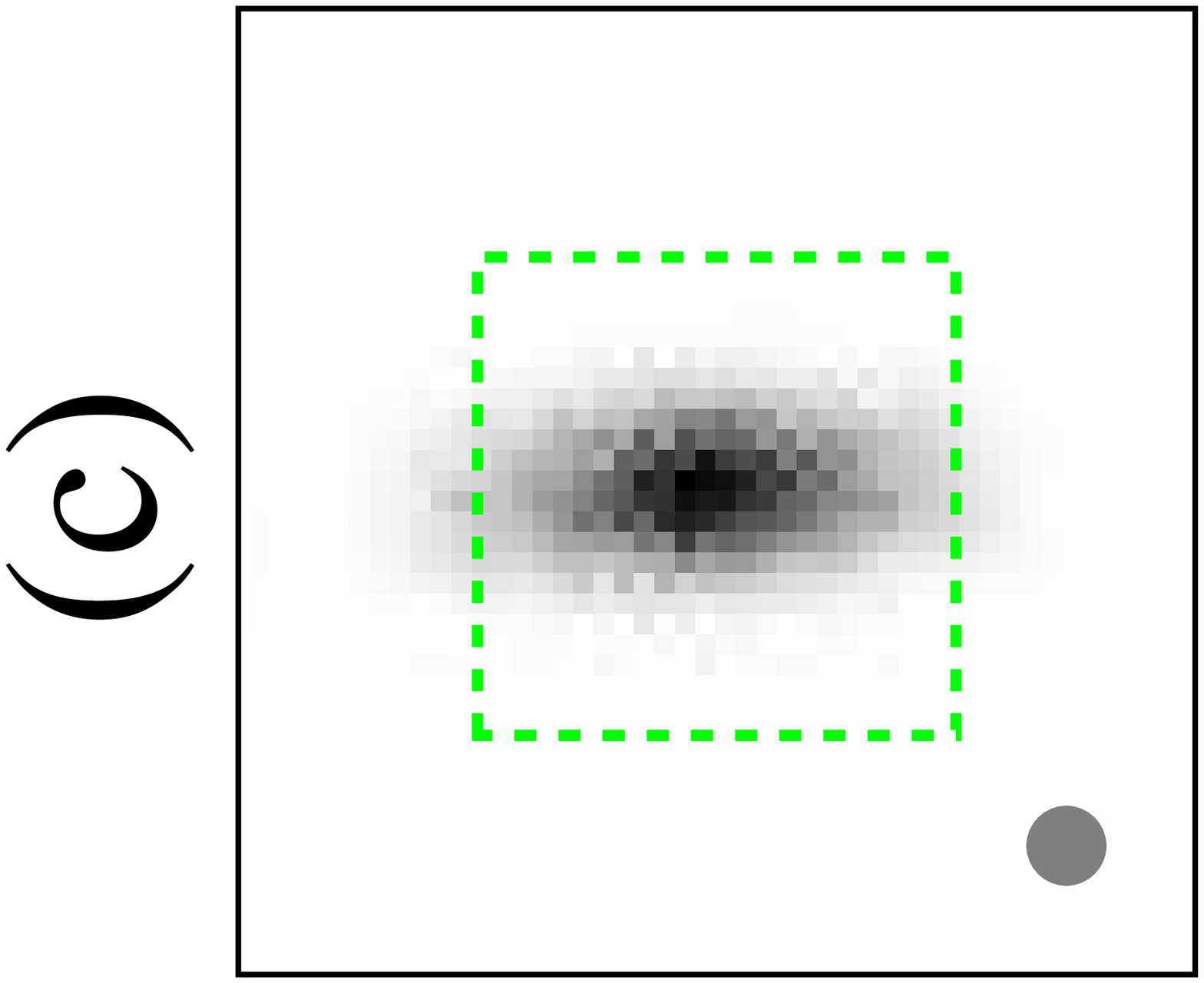} 
   \end{minipage}
   \begin{minipage}{0.4\textwidth}
    \psbox[r,xsize=0.25#1]{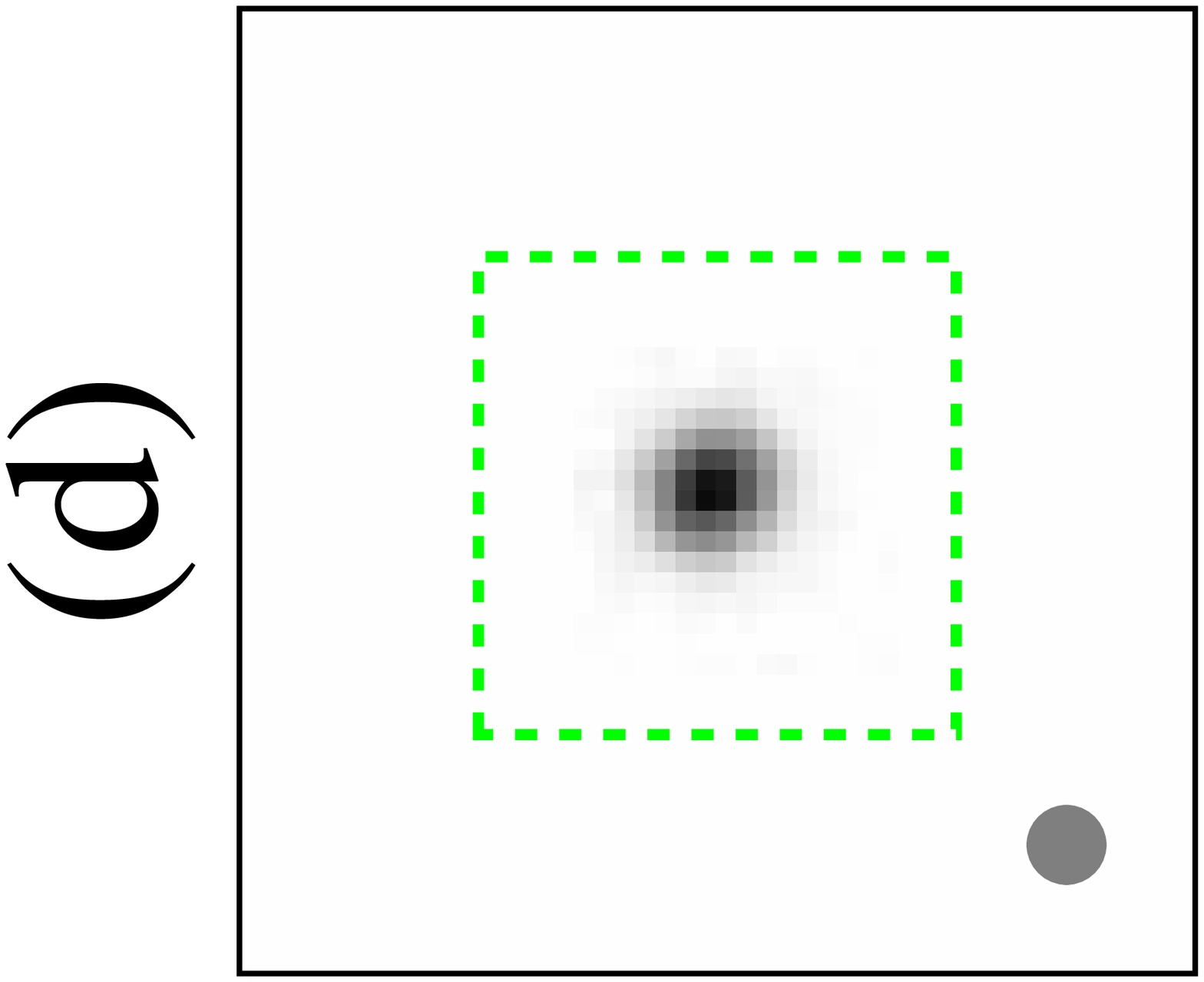} 
   \end{minipage}
 \caption{Position accuracies using our method shown in a $24\,\mu$m
 square.  The small circle in the lower left shows the shape of the mesh
 hole shadow on the CCD and the dashed square represents the size of the CCD
 pixel.  Position accuracy for single pixel events (a), for vertically
 split two-pixel events (b), for horizontally split two-pixel events (c)
 and for corner events (d).}
  \label{estimatePos}
   \end{center}
 \end{figure}

Figure~\ref{estimatePos} shows the distribution of {\bf $\Delta$} for
various event grades.  In this figure, we plotted the distribution of
{\bf $\Delta$} in the $24\,\mu$m square.  The mesh hole shadow is also
shown by the small circle on the lower left side of the figures and the
dashed squares 
represent the size of the CCD pixel for comparison.  We fitted the
results by two-dimensional Gaussian function and obtained the size of
the major and minor axes.  The major and minor axes always coincide with
the horizontal or vertical directions on the CCD.  It should be noted that
Fig.\,\ref{estimatePos} denotes the actual position accuracy convoluted with
the mesh hole shape.  The geometrical mesh hole size is $2.1\,\mu$m in
diameter while the effective mesh hole shadow on the CCD is expanded by
the diffraction.  The actual mesh hole shadow on the CCD is about
$2.2\,\mu$m for Ti-K X-rays in our configuration.  Hence we can eliminate
the effect of the mesh hole shape to evaluate the position accuracy of our
method summarized in Table~\ref{summary}.

For single pixel events, only one pixel contains a significant fraction of
the charge while the surrounding pixels contain noise.  Therefore, the
landing position inside the pixel is well within the pixel boundary
where $D$ is almost constant.  The actual shape of the region where the
single pixel event is generated is well within the pixel boundary. The
position uncertainty is slightly better than the pixel size.

The position accuracy can be improved when the charge splits into
adjacent pixels.  When the X-ray landing position is sufficiently close to the
vertical boundary, the X-ray event forms a horizontally split two-pixel
event.  In this case, the X-ray landing position can be improved along
the horizontal direction.  However, there is almost no improvement in
the vertical direction.  A similar improvement can be achieved for
a vertically split two-pixel event. We note that the position accuracy
of split direction for the vertically split two-pixel event is better
than that of the horizontally split two-pixel event.  The smaller extent
of the charge cloud has higher electron density than the bigger extent
of the charge cloud. The higher electron density will lead to a finer
position accuracy. Therefore, the above fact is consistent with the
elongation of the charge cloud shape~\cite{tsunemi99}.  Finally, when an
X-ray lands near the pixel corner, the charge splits both along the
horizontal  and  the vertical directions.  Then we can
improve the two-dimensional position resolution.

Yoshita et al.~\cite{yoshita99} have already demonstrated that the CCD
has a subpixel spatial resolution using split events.  They obtained
the position accuracy to 1.5 $\sim$ 2.0\,$\mu$m.  However, they employed
only two-pixel split events, resulting in the improvement of the position
resolution for one dimension.  This is mainly due to the experimental
setup: they employed a single pitch mesh experiment.  They had
difficulties in precisely determining the X-ray landing position for
corner events. However, we employed a multi-pitch mesh experiment by
which we could unequivocally determine the X-ray landing position for all
X-ray events. Hence, we could improve the two-dimensional position
resolution for the first time.  Furthermore, we introduced the mesh with
smaller holes to improve the position accuracy.

%%%%%%%%%%%%%%%%%%%% addeted at 2000/10/23 by j.hiraga %%%%%%%%%%%%%%%

\subsection{Position estimation using the event pixel}
\begin{figure}[hbtn]
%  \begin{center}
\hspace*{-30mm}
 \psbox[r,xsize=0.64#1]{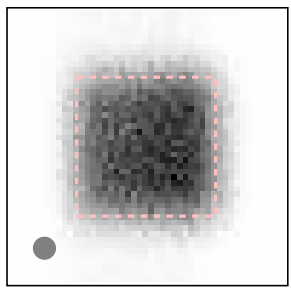} 
\vspace{-40mm}
   \caption{Same as for Fig.4 using the center of the event pixel as the X-ray landing position.}
   \label{conventional}
%  \end{center}
 \end{figure}

This is a very conventional method where the center of the event pixel
is considered to be the X-ray landing position. The position accuracy
using this method should be determined by the convolution between the
pixel shape and the mesh hole shadow. If we assume that the X-ray
detection efficiency is uniform over the pixel, we can calculate the
position accuracy which is 3.6\,$\mu$m (standard deviation, $\sigma$)
using $12\,\mu$m square pixels and $2.2\,\mu$m diameter mesh holes.  We
evaluated the accuracy of the position estimation employing this
conventional method.  Figure~\ref{conventional} shows the distribution
of {\bf $\Delta$} using all the X-ray events. We will analyze the result
by the same method used in our study.  The results listed in
Table~\ref{summary} are almost consistent with what we expected.  A
small difference between our results and those we expected is probably
due to the nonuniformity of the X-ray detection efficiency over the
pixel.

\vspace{-40mm}

\begin{table}
\caption{Accuracy of the incident X-ray position}
\label{summary}
 %\begin{center}
 \hspace*{-10mm}
\begin{tabular}{rccc} \hline
Ti-K X-rays                             & Branching ratio	& Horizontal
direction	& Vertical direction	\\
					&  (\%)			&  ($\mu$m)		&  ($\mu$m)		\\
 \hline
Single pixel event       		&  55.2		&  3.1$\pm$ 0.2		&   3.4$\pm$ 0.2		\\
Horizontally split two-pixel event	&   20.3		&  1.0$\pm$
0.1		&   3.1$\pm$ 0.2		\\
Vertically split two-pixel event	&   13.0		&  3.4$\pm$
0.2		&   0.6$\pm$ 0.1		\\
Corner event				&   11.5		&
 0.7$\pm$ 0.1		&   0.7$\pm$ 0.1		\\
\hline
Event pixel & & 3.4$\pm$ 0.2 & 3.9$\pm$ 0.2 \\
\hline
\end{tabular}
  %\end{center}
\end{table}

\section{Discussion}
So far, the spatial resolution of the CCD is considered to be
limited by the pixel size.  Furthermore, the charge cloud generated by
X-ray photons expands, resulting in the incident X-ray position blur.

By the conventional method, the position accuracy ($\sigma$) determined
by the pixel size would be approximately 0.3 times of the pixel size.
Tsunemi et al.~\cite{tsunemi98} reported a method to improve the
position accuracy up to 0.13 times of the pixel size.  They estimated
the X-ray landing position by employing the center of gravity of the
X-ray event pattern, particularly for the two-pixel split events.
However, the center of gravity of the event does not precisely show the
X-ray landing position since the charge cloud size is smaller than the
pixel size.  The precise X-ray landing position should be determined by
referring to the charge cloud shape.  Therefore, our method, considering
the charge cloud shape, can improve the accuracy of the X-ray landing
position better than that using the center of gravity.  Our result shows
the accuracy of the X-ray landing position up to 0.06 times the pixel
size, which shows that the CCD for X-ray use can be an image sensor with
a sub$\mu$m resolution.

The charge cloud size of the X-ray used in our experiment is relatively
small compared with the CCD pixel size employed.  This combination makes
the fraction of the single pixel event high and that of the split event,
particularly the corner event, low as described in
Table~\ref{summary}. The improvement of the position resolution can be
effectively attained when the charge splits into the adjacent pixel.
Since our method is practically useful only for corner events, we have
to increase the fraction of the split events in order to make the entire
area of the X-ray CCD as an image sensor with a sub$\mu$m resolution.
What are the practical conditions by which we can obtain the subpixel
resolution using our method more effectively?  There are two ways to do
this: one is to make the charge cloud size big and the other is to
manufacture a CCD chip with small pixel size.

The charge cloud shape depends on the travel distance between the
photoabsorption location and the potential well.  The longer the travel
distance, the bigger is the charge cloud shape.  Since the CCD employed
is a front illumination type (FI) CCD, the travel distance becomes
longer for higher X-ray energy, resulting in a bigger charge cloud
shape.  However, this does not apply to a back illumination type (BI) CCD.
In the BI CCD, the X-ray photon enters from behind the
depletion region.  X-rays with shorter photoabsorption length generate a
charge cloud at the far end of the depletion region.  The charge cloud
will travel a relatively long distance, resulting in a relatively large
cloud shape.  Therefore, a low energy X-ray in the BI CCD will generate
a larger charge cloud than that in the FI CCD.  Our method will function
efficiently in the BI CCD.

The accuracy of the X-ray landing position depends on the accuracy of
the charge cloud shape.  The actual charge cloud shape will depend on
various parameters; the incident X-ray energy, the thickness of the
depletion region and the working condition of the CCD, and the applied
voltage on the gates and the clocking pattern.  We can directly measure the
charge cloud shape using the mesh experiment.  However, obtained the
charge cloud shape is a mean shape for a given X-ray energy, it will be
different for individual X-ray photons depending on the photoabsorption
depth.  This will become a position error through our method.

Our method requires the charge cloud generated by the X-ray photon to
spill over the adjacent pixel.  Therefore, we need not make the CCD
pixel very small if the pixel size is comparable to the primary charge
cloud size.  Furthermore, we need a CCD with a thicker depletion region to
obtain a higher detection efficiency for high energy X-ray.  We only
require the pixel size to be small enough to generate split events.  What types
 of grades generated by an X-ray photon depend not only on where the
X-ray lands but also on how deep the X-ray is photoabsorbed in the
depletion layer which is a stochastic process.  The charge cloud size
($\sigma$) measured for Ti-K(4.5\,keV) X-rays is 1.0\,$\sim$ 1.4\,$\mu$m.
Considering the noise level, we can detect the split charge to the
adjacent pixel when the X-ray photon lands within about $3\,\mu$m away
from the boundary.  Therefore if the pixel size is about $6\,\mu$m
square, we can obtain split events wherever the X-ray lands.  In the
present CCD, having $12\,\mu$m square pixel, X-rays landing on the
central part of the pixel become single pixel events even if they are
photoabsorbed deep in the depletion region.

We should note that the X-ray photoabsorbed in the shallow region
generates a very small charge cloud size in the FI CCD.  Therefore, some
fraction of X-ray photons generates single pixel event however small the
pixel size is.  Whereas no X-ray photon landing onto the central part of
the pixel generates a split pixel event if the pixel size is bigger than the
charge extent.  The branching ratio of the X-ray events does not
represent the area generating each event grade but represents what
 fraction of the X-ray events becomes what type of grades.  Hence, we
 can expect that some X-ray events will generate split events
even if they land at the center of the pixel of the CCD with about
6\,$\mu$m square pixel.  Since the CCD chip employed at present has
a 12-$\mu$m-square pixel, we can say that the CCD with two times smaller
pixel will be an image sensor with a sub$\mu$m resolution.

\section{Conclusion}

We performed the multi-pitch mesh experiment using the gold mesh of
$2.1\,\mu$m diameter holes.  Then, we calculated the function,
$D$, which shows the output from the event pixel as a function of the
X-ray landing position.  The primary charge cloud shape generated by
X-ray photons can be measured from $D$.

Using the charge cloud shape, we can calculate the X-ray landing
position with subpixel resolution.  Our method compares the
$3\times 3$ pixel data with $D$.  The position resolution is improved in
the direction where the charge splits.  Therefore, there is almost no
improvement for single pixel events.  The position resolution for
two-pixel split events can be improved in one direction.  When the X-ray
photon lands near the pixel corner, the charge splits both in the
horizontal direction and the vertical direction generating corner
events.  Then the two-dimensional position resolution can be improved to
subpixel resolution.  The accuracy of the position is less than
$1\,\mu$m in the direction where the charge splits.

We have discussed how we can improve the position resolution of the CCD
for X-ray use by introducing our method.  Our method becomes practical
when an X-ray photon produces a split event.  For the optical region, a
CCD with very small pixel size is manufactured whereas it is not
designed for X-ray use.  Taking into account the noise level in our
system, we find that the CCD with a pixel size of about $6\,\mu$m square will
generate split events from the entire region.  Therefore, the CCD can be
used as an image sensor with a sub$\mu$m resolution for X-ray photons.

%\acknowledgement
\section{Acknowledgements}

The authors are grateful to all the members of the CCD team in Osaka
University.  This work is supported by the ACT-JST program, Japan Science
and Technology Corporation and the SUMITOMO FOUNDATION.  J.H. was
partially supported by JSPS Research Fellowship for Young Scientists,
Japan.

\end{document}